\documentclass[5p,times]{elsarticle}

\usepackage{amssymb}
\usepackage{amsmath}
\usepackage{listings}

\usepackage{enumitem}
\usepackage{microtype}
\usepackage{algorithm}
\usepackage{subcaption}
\usepackage{algorithmic}
\usepackage{tcolorbox}
\usepackage{graphicx}
\usepackage{float}
\usepackage{fancyhdr}
\usepackage{array}
\usepackage{stfloats}
\tcbset{
  left=2mm, %
  right=2mm, %
  top=1mm, %
  bottom=1mm %
}

\journal{Journal of Systems and Software
}

\begin{document}

\begin{frontmatter}

\title{Atys: An Efficient Profiling Framework for Identifying Hotspot
Functions in Large-scale Cloud Microservices} 

\author[SJTU]{Jiaqi Sun}
\ead{jotaro@sjtu.edu.cn}
\author[ZJU]{Dingyu Yang}
\ead{yangdingyu@zju.edu.cn}
\author[SJTU,KeyLab]{Shiyou Qian\corref{corresponding}}
\ead{qshiyou@sjtu.edu.cn}
\author[SJTU,KeyLab]{Jian Cao}
\ead{cao-jian@sjtu.edu.cn}
\author[SJTU,KeyLab]{Guangtao Xue}
\ead{gt\_xue@sjtu.edu.cn}

\cortext[corresponding]{Corresponding author.}
\affiliation[SJTU]{organization={Department of Computer Science, Shanghai Jiao Tong University},
            addressline={800 Dongchuan RD}, 
            city={Shanghai},
            postcode={200240}, 
            state={Shanghai},
            country={China}}
\affiliation[KeyLab]{organization={Shanghai Key Laboratory of Trusted Data Circulation and Governance in Web3},
            addressline={800 Dongchuan RD}, 
            city={Shanghai},
            postcode={200240}, 
            state={Shanghai},
            country={China}}

\affiliation[ZJU]{organization={The State Key Laboratory of
Blockchain and Data Security, Zhejiang University},
            addressline={38 Zheda RD}, 
            city={Hangzhou},
            postcode={310027}, 
            state={Zhejiang},
            country={China}}

\begin{abstract}
To handle the high volume of requests, large-scale services are comprised of thousands of instances deployed in clouds. These services utilize diverse programming languages and are distributed across various nodes as encapsulated containers. Given their vast scale, even minor performance enhancements can lead to significant cost reductions. In this paper, we introduce Atys\footnote{Atys, known as the god of tracking and discovery, is often portrayed as a skilled hunter and scout in mythology.}, an efficient profiling framework specifically designed to identify hotspot functions within large-scale distributed services. 
Atys presents four key features.
First, it implements a language-agnostic adaptation mechanism for multilingual microservices. Second, a two-level aggregation method is introduced to provide a comprehensive overview of framegraphs. Third, we propose a function selective pruning (FSP) strategy to enhance the efficiency of aggregating profiling results.
Finally, we develop a frequency dynamic adjustment (FDA) scheme that  dynamically modifies sampling frequency based on service status, effectively minimizing profiling cost while maintaining accuracy.
Cluster-scale experiments on two benchmarks show that the FSP strategy achieves a 6.8\% reduction in time with a mere 0.58\% mean average percentage error (MAPE) in stack traces aggregation. Additionally, the FDA scheme ensures that the mean squared error (MSE) remains on par with that at high sampling rates, while achieving an 87.6\% reduction in cost.
\end{abstract}

\begin{keyword}
Data center\sep performance profiling\sep distributed system\sep cloud computing
\end{keyword}

\end{frontmatter}

\section{Introduction}
\label{sec:intro}

Large-scale online services like Google search \citep{googlesearch} and Amazon \citep{amazon} play a vital role in our daily life. Typically  deployed in clouds, these services ensure continuous availability. To manage the substantial influx of requests, they rely on numerous instances(or replicas). For example, a one-day trace in Meta's production environment on 2022/12/21 shows its microservice topology contained 18,500 active services and over 12 million service instances. The frontend service \textit{\small{www}} is the most replicated service with as many as 557,000 instances, as it handles most incoming requests\citep{meta-trace}.

Identifying and optimizing hotspot functions in large-scale cloud services can significantly improve resource efficiency. These hotspot functions, which include both business logic and infrastructure components, account for a significant portion of service resource consumption. Due to their vast scale, even minor performance enhancements can lead to substantial cost savings. 

Large-scale services markedly differ from traditional monolithic applications in hardware and software architectures, as well as deployment strategies. Hardware-wise, these services are spread across numerous worker nodes in clouds \citep{DBLP-series-synthesis-2018Barroso}. On the software front, they employ various programming languages, encompassing both compiled and interpreted types. The rise of container technology has transitioned deployment methods in clouds from virtual machines to containers \citep{Merkel2014DockerLL}. 
Administrators frequently employ performance profiling tools to identify hotspot functions.

Current profiling tools can be divided into single-machine and multi-machine profilers, both of which face challenges in profiling large-scale distributed services. 
Single-machine profilers are often language-specific. For instance, system-level tools like Perf \citep{Melo2010TheNL} can only collect native symbols, preventing them from recognizing user-defined functions and file names in interpreted languages like Python.  
Additionally, profilers such as Intel VTune \citep{vtune} and gprof \citep{10.1145/872726.806987} are not suitable for large-scale distributed environments. On the other hand, multi-machine profilers, while leveraging single-machine profilers for individual node call stack tracing, struggle with access restrictions and usability issues.
Google Wide Profiling (GWP) \citep{5551002}  is restricted to internal users, while HPCToolkit \citep{DBLP:journals/concurrency/AdhiantoBFKMMT10} necessitates specific commands for application initialization, limiting direct process attachment.
Additionally, constructing multi-machine profilers on monitoring platforms, as shown by Pyroscope \citep{pyroscope}, is achievable. However, indiscriminate use of single-machine profilers in multi-instance services can lead to considerable performance overhead.

In this paper, we present a distributed profiling framework named \textbf{Atys}, which leverages existing profiling tools to identify hotspot functions in a cost-effective and non-invasive way. Our design principle focuses on the unique characteristics of large-scale microservices, emphasizing both cost efficiency and ease of use.

Firstly, \textbf{Atys} enables hotspot function profiling for both compiled and interpreted languages within containers by utilizing specialized profiling tools. Secondly, to tackle the issue of visualizing call stacks across multiple instances of the target service, \textbf{Atys} aggregates flamegraphs \citep{flamegraph} from various instances to pinpoint hotspot business logic functions, going beyond library functions. Thirdly, acknowledging the multi-threaded nature of highly concurrent microservices, \textbf{Atys} implements a function selective pruning (FSP) strategy that consolidates profiling results at the thread level, facilitating cost-effective profiling. Lastly, \textbf{Atys} introduces a frequency dynamic adjustment (FDA) scheme that adaptively adjusts sampling rate based on profiling outcomes, significantly lowering profiling costs while minimizing accuracy loss.

To evaluate the effectiveness of \textbf{Atys}, we conducted experiments on a cluster consisting of 13 nodes, using typical Java and Python workloads Specjbb2015 \citep{specjbb} and VGG16 \citep{Simonyan2014VeryDC} as benchmarks. The experiment results show that: 
\begin{itemize}
    \item The FSP strategy reduces aggregation time by 6.8\% while retaining the 99th percentile (P99) hotspot threads, resulting in a mean absolute percentage error (MAPE) of just 0.58\% for the top 50 hotspot functions.
    \item The FDA scheme, averaging a sampling rate of 3,578 Hz, yields results with a mean squared error (MSE) comparable to a 10,000 Hz sampling rate, at only 12.4\% of the cost.
    \item We demonstrate the scalability of \textbf{Atys}; when sampling 1,000 service instances concurrently, the central server's average memory usage is merely 175 MB, with overall CPU usage remaining below 6\%.
\end{itemize}
Overall, \textbf{Atys} demonstrates significant versatility, accuracy, and cost-effectiveness in a non-invasive manner within containerized and distributed environments, meeting the requirements of modern large-scale cloud services.

The main contributions of this paper are summarized as follows:
\begin{itemize}
    \item We present an efficient profiling framework called \textbf{Atys}, specifically designed for large-scale microservices in production data centers.
    
    \item We propose a language-agnostic mechanism that accommodates both compiled and interpreted programming languages, alongside a two-level flamegraph aggregation method that offers a comprehensive overview of multiple service instances.
    
    \item We propose an efficient function selective pruning strategy along with an adaptive scheme for adjusting sampling frequency, significantly reducing profiling costs.
    
    \item We implement a prototype\footnote{The code of \textbf{Atys} can be accessed  at: https://github.com/ottoSJTU/Atys} of \textbf{Atys} and evaluate its performance through a series of experiments.
\end{itemize}

\section{Related Work}

In this section, we review the existing literature from three perspectives: profiling paradigms, profiling tools, and profiling optimization techniques.

\subsection{Profiling Paradigms}
Performance profiling involves analyzing a program's runtime behavior by capturing its hardware or software metrics, essential for identifying performance bottlenecks. Profilers generally rely on sampling for improved efficiency or code instrumentation for enhanced precision.

\subsubsection{Sampling}
Sampling-based profiling, or stack tracing, captures periodic snapshots of a program's function call stacks. The last function in these traces represents the active function during sampling. As samples accumulate, the frequency of each function's appearance reveals its resource consumption over the observed time window.

Recent studies indicate that this paradigm may exhibit slight inaccuracies \citep{DBLP:conf/ics/XuWSJ019} \citep{DBLP:conf/apsys/YiDDC20}. Xu et al. identify that the function represented in the sample may not always correspond to the executed one, a phenomenon known as the \textit{skid effect} \citep{DBLP:conf/ics/XuWSJ019}. Similarly, Intel's \textit{Precise Event Based Sampling (PEBS)} also experiences a related \textit{shadow effect} \citep{DBLP:conf/apsys/YiDDC20}.

However, sampling incurs minimal CPU overhead, as it interrupts the program only upon receiving specific signals. Furthermore, it is deemed non-invasive since it does not necessitate modifications to the profiled program. Consequently, its efficiency contributes to its popularity in major profiling tools such as Perf \citep{Melo2010TheNL}, HPCtoolkit \citep{hpctkt-wf}, and Intel VTune \citep{vtune}.

\subsubsection{Code Instrumentation}
Code instrumentation entails integrating supplementary code into the target program. The modified program generates performance data during execution, which is subsequently utilized for report generation.

This paradigm ensures accuracy by collecting metrics solely when relevant functions are active, thereby mitigating issues like the shadow effect \citep{DBLP:conf/apsys/YiDDC20} and skid effect \citep{DBLP:conf/ics/XuWSJ019} identified in recent studies. However, this technique often requires program restarts or recompilation, leading to significant overhead. Furthermore, the additional instructions from profiling code can degrade service response times and increase resource consumption \citep{10.1145/3679007.3685058, 10.5555/243846.243857}. 

Modifying source code, recompiling, and restarting services are impractical in large-scale microservice environments. Therefore, non-invasive sampling-based profiling is favored in this scenario for its minimal impact on service operations.

\subsection{Profiling Tools}
Current profilers are categorized into two main types: single-machine profilers and multi-machine profilers, with the latter typically built on the former's foundation. Single-machine profilers operate on individual worker nodes to trace the call stack of a process or application. However, they are insufficient for profiling multiple instances across distributed worker nodes. Consequently, researchers integrate single-machine profilers with data links, databases, dashboards, and other components into multi-machine profilers to effectively capture large-scale applications and manage profiling metrics cohesively.

\subsubsection{Single-machine Profilers}
When analyzing application performance on a single worder node, the Linux-based tool Perf \citep{Melo2010TheNL} is widely utilized. However, Perf does not support all interpreted languages. Fortunately, most of these languages have built-in profilers, such as \textit{pprof} for Golang \citep{pprof}, \textit{cProfile} for Python \citep{cprof}, and \textit{V8 profiler} for Node.js \citep{v8prof}. 
These profilers are tailored for their specific languages, enabling precise metric collection at the line level and directly linking metrics to their source. Although built-in profilers provide excellent compatibility, they necessitate modifications to the source code, which can pose challenges.

To address this issue, external profilers have been created for specific programming languages. Notably, \textit{Async-profiler} \citep{asyncprof} and \textit{JProfiler} \citep{jprof} are widely utilized for Java. These profilers operate as Java agents, leveraging the Java Instrumentation API \citep{javainstru}. They can be easily launched or attached to the Java Virtual Machine (JVM), enabling stack trace sampling on demand.
These tools are not only user-friendly but also provide language-specific enhancements. For instance, Async-profiler employs the \textit{AsyncGetCallTrace} API from OpenJDK \citep{asygetct} to mitigate the prevalent safepoint bias issue in Java profiling.
In the realm of large-scale distributed services, these agents are particularly advantageous as they can be seamlessly integrated into individual nodes with minimal disruption to the host system.

\subsubsection{Multi-machine Profilers}
Currently, several multi-machine profilers are available for developers. Notably, Google's continuous profiling infrastructure for data centers, known as Google-Wide Profiling (GWP) \citep{5551002}, offers high scalability and the ability to attribute metrics to source code using symbols. However, GWP is exclusive to Google, with limited public information accessible. Other proprietary multi-machine profilers include Prodfiler \citep{prodfiler} and Google Cloud Profiler \citep{googlecloudprof}, both of which offer limited functionality in their free versions.

Developers can create custom monitoring systems as multi-machine profilers. For example, Performance Co-Pilot (PCP) \citep{pcp} and Zabbix \citep{zabbix} use ``agents" to gather various performance metrics from individual targets. In Prometheus \citep{prometheus}, this is termed as ``exporter." 
Regardless of terminology, these "agents" or "exporters" primarily function as single-machine profilers. Pyroscope \citep{pyroscope} employs a similar approach, utilizing the same profiling backend as \textbf{Atys} (async-profiler for Java and py-spy for Python). However, directly deploying these kernels without optimizations leads to significant performance degradation, as discussed in Section \ref{eval}. In \textbf{Atys}, we implement a comparable architecture while optimizing to reduce profiling overhead compared to traditional single-machine profilers.

\subsection{Profiling Optimization Techniques}
In addition to engineering methods, current research focuses on enhancing profiling accuracy while minimizing costs, which aligns with our objectives. From a bottom-up perspective on profiling tools, the relevant levels include microarchitecture, event, sampling strategy, and system. Existing work primarily optimizes around these levels.

Optimizations of single-machine profilers focus on microarchitecture, event, and sampling strategy levels.

At the microarchitecture level, current researches are dedicated to accurately obtaining the cost consumed by each CPU instruction. Gottschall et al. demonstrate that each static CPU microinstruction impacts overall execution time differently, leading to the development of time-proportional Per-Instruction Cycle Stacks (PICS) \citep{10.1145/3295500.3356167}. Recognizing that long-latency memory accesses in data-intensive applications (epochs) can be overlapped with useful computations or with one another, Carlson et al. propose an epoch-based microarchitectural analysis to model the performance effects of microarchitecture interactions with these applications \citep{7128801}. Moreira et al. find that analyzing various instruction types within a block yields a broader range of insights. They introduce the Block-Level Architecture Profiler (BLAP), an online mechanism that operates at the basic block level while profiling microarchitectural bottlenecks, such as delinquent memory loads, hard-to-predict branches, and contention for functional units \citep{6970668}.
    
At the event level, a significant challenge is the disparity between the limited number of metric registers and the vast array of target events. Modern microprocessors can capture only a few hardware metrics simultaneously, which is inadequate given the number of target events. To address this issue, researchers have proposed time interpolation techniques that allow for the collection of different metrics at various time points, subsequently interpolating the results to facilitate reasoning across all metrics at the same time points \citep{4408263}. Related approaches include multiplexing \citep{10.1145/1088149.1088163, 1521115} and trace alignment \citep{10.1145/1094811.1094834}. Mytkowicz et al. assess how well these techniques maintain time-varying relationships between metrics, providing insights into their reliability \citep{4408263}.
    
At the sampling strategy level, researchers focus on dynamically optimizing the sampling process based on current results during system operation, rather than following fixed parameters beforehand. Ehlers et al. utilize time series analysis of operational response times alongside the architectural information of the target system to facilitate rule-based monitoring adaptations for sampling coverage \citep{10.1145/1998582.1998628}. Cho et al. introduce a sampling-based code instrumentation technique known as \textit{instant profiling} \citep{instantprof}. Rather than directly sampling stack traces or instrumenting the entire execution, they periodically interleave native and instrumented executions, effectively determining which code segments to instrument through this sampling strategy.

On the other side, academic research on multi-machine profilers primarily addresses system-level challenges. In distributed profiling systems, comparing timestamps across different machines is essential, as the relative behavior of system components is captured by timestamping significant events on each machine \citep{5600322}. 
Benavides et al. introduce FreeZer, a timestamp synchronization algorithm that effectively bounds timestamp inaccuracies within a specified time interval by analyzing the relative frequency and zero bounds between clock pairs \citep{10.1145/3360582}. 
To tackle the timestamp synchronization issue in big data processing environments, Cloudprofiler has been developed to measure event durations within a streaming framework. It leverages time-stamp counters (TSCs) during network quiescent periods to achieve accuracy within tens of microseconds \citep{yang2023cloudprofilertscbasedinternodeprofiling}.
Another objective is to extend the optimization of single machine profilers to the system level. Ahn et al. introduce SCOZ to shift the target of virtual speedup \citep{10.1145/3205911} from threads to CPU cores, thereby adapting causal profiling techniques from multithreaded applications to multicore systems \citep{https://doi.org/10.1002/spe.2930}.

In large-scale microservice environments, efficiency is paramount. Thus, \textbf{Atys} distinguishes itself by optimizing the sampling strategy to attain high efficiency while maintaining an acceptable level of accuracy.

\section{Design of Atys}

\subsection{\textbf{System Overview}}

\subsubsection{\textbf{Overall Architecture}}
\begin{figure}[t]
\centering
\includegraphics[width=0.5\textwidth]{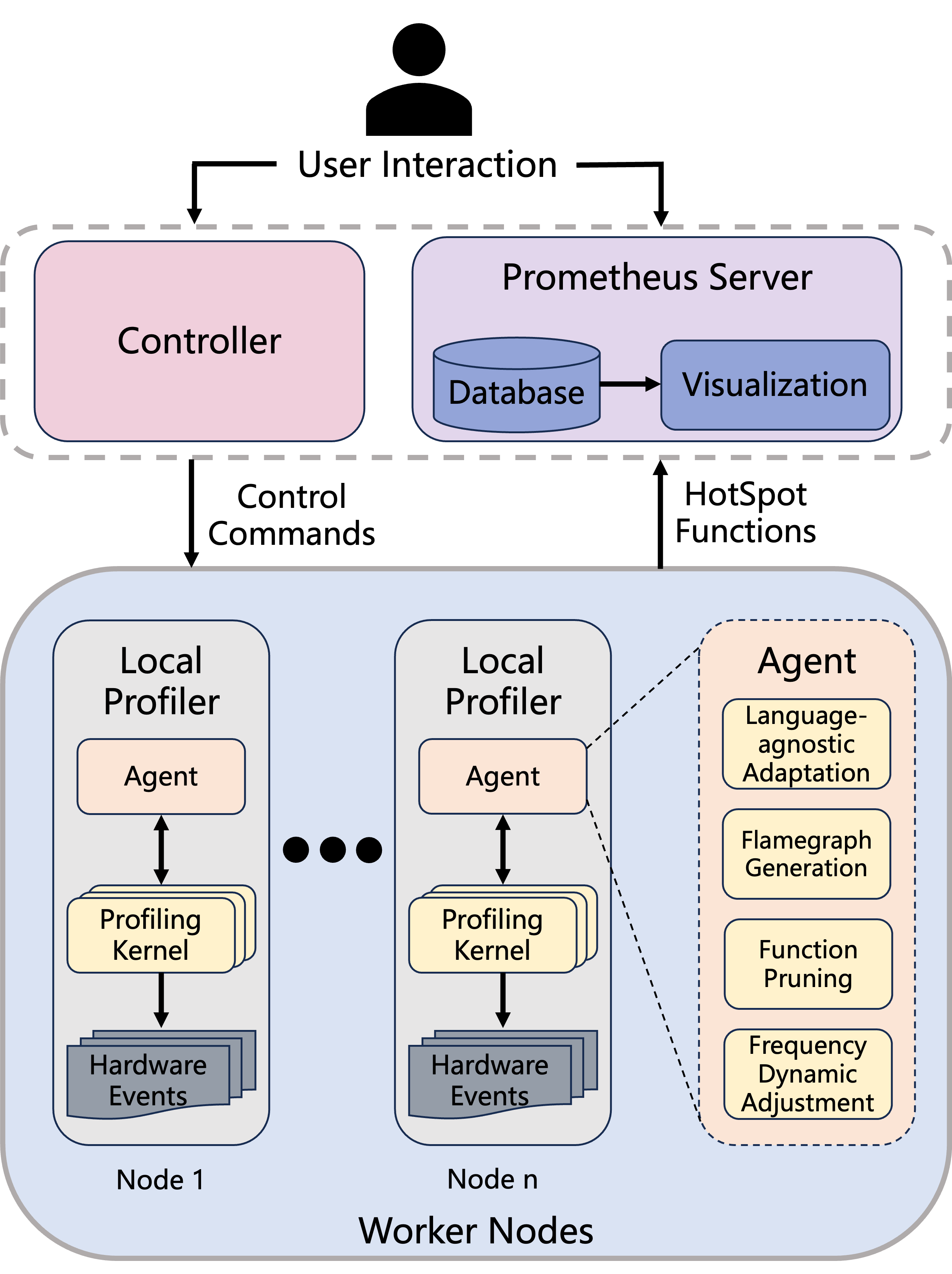}
\vspace{-3mm}
\caption{The architecture of \textbf{Atys}}
\label{system}
\end{figure}

The architecture of \textbf{Atys} is depicted in Figure \ref{system}, consisting of three key components. 1) \textbf{Local Profiler:} This component samples and processes stack traces from the target microservice, subsequently exposing metrics that indicate hotspot functions to the Prometheus server. 
2) \textbf{Controller:} Acting as a mediator, the controller receives user commands and manages local profilers. 
3) \textbf{Prometheus Server:} This server pulls metrics exposed by local profilers and provides services for data storage, analysis, and visualization.

Prometheus \citep{prometheus} is an open-source monitoring system that has gained significant traction in monitoring large-scale distributed systems. In \textbf{Atys}, the persistent data storage and data transmission link are built on Prometheus. Metrics generated in profiled nodes are retrieved via the HTTP protocol and stored in a built-in time series database (TSDB), which allows for efficient snapshots and flexible data slicing. Users can leverage the flexible query language PromQL to easily query, analyze stored metrics, or visualize them on the Prometheus dashboard.

The design principle of \textbf{Atys} is to accommodate the distributed architecture and varied workloads of large-scale microservices. This entails deploying local profilers equipped with multiple profiling kernels on each worker node while establishing central nodes for unified control and data processing.
This design presents two key advantages. First, the workload of \textbf{Atys} is mainly focused on the Prometheus server side. By ensuring that local profilers utilize minimal CPU time, this design reduces interference with the target microservices. Second, the functionalities of various components are loosely coupled, which improves the system's maintainability and scalability.

\subsubsection{\textbf{System Workflow}}
The workflow below outlines how \textbf{Atys} initiates a profiling task:
\begin{enumerate}
    \item The user creates a configuration file that specifies the target node, target service, metrics to be sampled, sampling frequency, and other required parameters.
    \item The controller parses the configuration file, connects to the target nodes, and initiates profiling tasks while supplying necessary parameters to the local profilers.
    \item The local profilers perform continuous sampling of stack traces. 
    \item The local profilers aggregate the raw stack traces collected during profiling and compute certain metrics that cannot be directly sampled. They then convert the profiling results into formatted Prometheus metrics \citep{prometheusMetric} and upload them to an available port.
    \item The Prometheus server monitors the ports exposed by the local profilers, pulls the metrics, and stores them in its built-in time-series database. 
    \item The user can conveniently access visualized data on the intuitive Prometheus dashboard or export it from the database for additional analysis.
\end{enumerate}

\subsection{\textbf{Local Profiler}}
A local profiler primarily identifies hotspot functions on individual worker nodes. It consists of two components: \textbf{profiling kernels} and the \textbf{agent}.

\subsubsection{Profiling Kernels}
\textbf{Atys} provides efficient and non-invasive profiling across various programming languages in containerized environments. It employs specialized ``\textbf{profiling kernels}" on worker nodes to execute stack tracing without altering or disrupting the target service. These kernels leverage the hardware Performance Monitoring Unit (PMU) to gather runtime stack traces triggered by specific events. 

Each stack trace records the sequence of function calls leading up to the sampling point. For example, in a trace such as $Func1 \rightarrow Func2 \rightarrow Func3$, \textit{Func3} represents the currently executing function, called by \textit{Func2}, which was in turn called by \textit{Func1}. Identical stack traces are counted to indicate their frequency of occurrence.

The selected profiling kernels for \textbf{Atys} must adhere to specific criteria. They should not necessitate program modifications or restarts, should be sampling-based, and must offer comprehensive support for the target language. In line with these requirements, async-profiler \citep{asyncprof} is chosen for Java, py-spy \citep{pyspy} for Python, and Perf \citep{Melo2010TheNL} for compiled languages.

\subsubsection{Agent}
The agent acts as a mediator connecting three core components of the system. It initially responds to the controller's instructions by automatically activating the appropriate kernels with dynamic frequencies. Subsequently, it processes the raw stack traces from these kernels, utilizing a function pruning strategy to minimize data aggregation costs. Finally, it converts the aggregated stack traces into Prometheus-compatible metrics, ensuring accessibility to the Prometheus server. Additionally, it regularly generates flame graphs for global aggregation.

\begin{figure}[!t]
\centering
\includegraphics[width=0.4\textwidth]{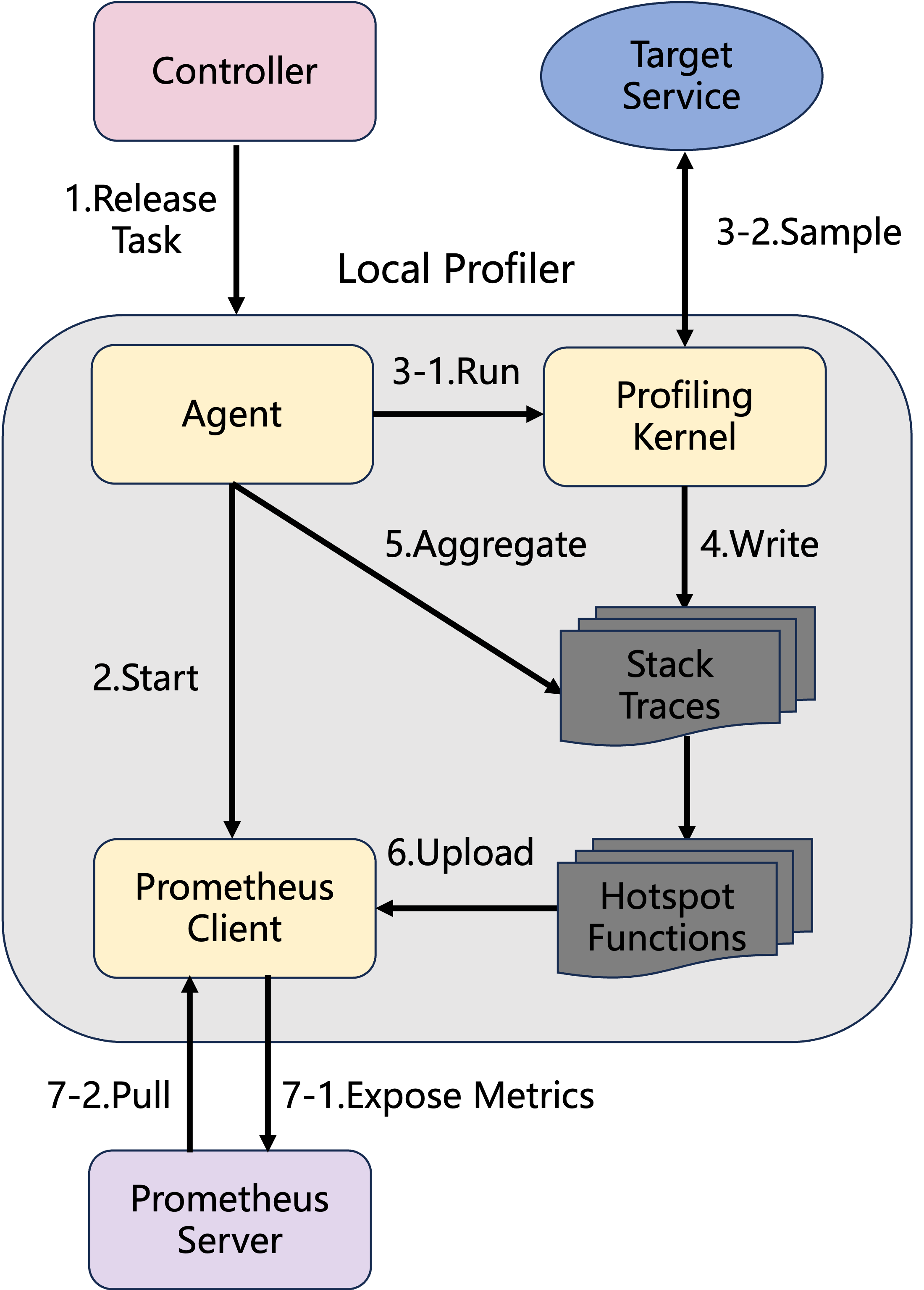}
\caption{Internal interactions in the Local Profiler}
\vspace{-3mm}
\label{local profiler}
\end{figure}

\subsubsection{Local Profiler Workflow}
As illustrated in Figure \ref{local profiler}, the local profiler functions as follows, supported by the two previously mentioned components: 
\begin{enumerate}
    \item  The agent receives commands from the controller that defines the profiling configurations.
    \item  The agent initiates a Prometheus client, setting up Prometheus metrics.
    \item  The profiling kernel is launched to sample stack traces, following preparatory steps that include obtaining necessary authorizations and embedding dynamic link libraries into the target container.
    \item  The profiling kernel consistently records raw stack traces.
    \item  The agent aggregates raw call stacks to identify hotspot functions within the target service and computes certain metrics that cannot be directly sampled, such as CPU time for functions based on sampling frequency and the number of samples per function.
    \item  The agent uploads the aggregated stack traces to the initialized Prometheus metrics.
    \item  The agent utilizes the Prometheus client to expose metrics via an available port.
\end{enumerate}

\subsection{\textbf{Controller}} 
The controller in \textbf{Atys} serves as the central hub for routing user commands to the system's components. Its primary function is to maintain a persistent waiting state, ready to receive commands via the Command Line Interface (CLI). Upon a user triggering profiling initialization, the controller interprets the configuration, establishes communication with local profilers, and assigns profiling tasks based on the user's specifications. Furthermore, the controller allows users to manage existing profiling tasks with ease.

\subsection{\textbf{Prometheus Server}}
The Prometheus server in \textbf{Atys} is essential for collecting metrics from the monitored ports of worker nodes. These metrics, reflecting the performance of hotspot functions, are generated by local profilers from raw stack traces and formatted according to Prometheus standards, which include metric names and a set of descriptive labels \citep{prometheusMetric}. 

In managing hotspot functions, the server employs a straightforward yet effective strategy: it adds a function to the trace if it has not been previously recorded or updates the existing trace with new resource consumption data from the latest profiling. Consequently, the Prometheus server offers a comprehensive data management suite for the tracked hotspots, encompassing persistent storage, querying, analysis, and visualization capabilities. This allows users to monitor resource consumption trends or export data for in-depth evaluation.

\section{Functionality Expansion}
To enhance usability, our goal is to reduce the prerequisite knowledge required for profiling large-scale services. Additionally, we strive to offer users a holistic view for identifying hotspot functions. 
To achieve this, we improve \textbf{Atys} in two key ways: by integrating a language-agnostic adaptation mechanism and by enabling the aggregation of multi-instance flamegraphs.

\subsection{Language-agnostic Adaptation Mechanism}
In modern application development, the use of various programming languages is prevalent, especially for large-scale services in data centers. These services often comprise multiple microservices, each potentially developed in a different programming language. 
According to Meta, the endorsed programming languages for server-side use include 16 types, with Hack, C++, Rust, and Python being the primarily supported, while others like Java and Go are supported in specific cases \citep{meta-trace, meta_server_side}. These languages include both compiled ones and interpreted ones.
However, current profiling tools face limitations in this context. Built-in profilers differ across programming languages, and even multi-machine profilers necessitate users to be knowledgeable about the language employed in the target service to choose an appropriate profiler. This requirement for language-specific expertise can be burdensome for users, as it is unrealistic to expect them to have a comprehensive understanding of each microservice that constitutes the service.

\textbf{Atys} presents a language-agnostic adaptation mechanism for multilingual microservices. It classifies programming languages into two categories: interpreted and compiled. For compiled languages, where binary code executes directly on the CPU, stack traces can be collected by directly accessing the service's stack space. In contrast, interpreted languages require reconstructing memory through interpreter-provided APIs due to their language-specific memory models. Ultimately, \textbf{Atys} automates the mapping of each programming language to its respective profiling kernel, significantly improving the convenience and efficiency of multilingual microservice support.


\textbf{Atys} initiates its process by verifying the presence of active interpreters associated with the target service. Upon identifying an interpreter, \textbf{Atys} selects the appropriate profiling kernel for the specific language, such as async-profiler for Java or py-spy for Python, to capture stack traces. 
If no interpreter is found, indicating that the target service is built with a compiled language, \textbf{Atys} seamlessly transitions to a system-level profiler like Perf to gather stack traces. This adaptable approach enables \textbf{Atys} to effectively profile services across multiple languages, ensuring a comprehensive user experience free from language constraints.


\subsection{Two-level Flamegraph Aggregation Method}
As described before, Meta's data center employs hundreds of thousands of instances for a single service, with a maximum count of 557,000 and the $99^{th}$ percentile is 31,306 \citep{meta-trace}. 
Given that large-scale services are inherently distributed with dynamically varying performance, focusing solely on resource-intensive functions may lead to an incomplete understanding of performance bottlenecks. To address this limitation, \textbf{Atys} employs flamegraphs alongside numeric metrics to accurately identify the true hotspot functions. While traditional tools can generate flamegraphs for individual instances \citep{flamegraph}, \textbf{Atys} enhances its capability by aggregating these across all instances, offering a comprehensive perspective.

Profiling tools rank functions by resource consumption, but this can be misleading. High resource usage may stem from low-level library functions that perform critical tasks like locking or random number generation. 
Crucially, the high resource usage of these library functions does not imply they are the primary targets for optimization. A more thorough analysis of the business logic functions is often required, as they may excessively invoke resource-intensive low-level library functions, revealing inefficiencies in the code logic. Thus, it is essential to scrutinize the entire function call stack rather than solely concentrating on the resource usage of low-level library functions.

Consider a Java program \textit{RandomGen} as an example. It executes the methods \textit{generate\_UUID\_100times()} and \textit{generate\_UUID\_300times()} to produce Universally Unique Identifiers (UUIDs) 100 and 300 times respectively, utilizing the \textit{UUID.randomUUID()} method from the standard library \textit{java.util.UUID}. This method employs multiple hash functions to generate a unique identifier.
In this program, users should focus on identifying potential issues in the business logic that may lead to excessive library function usage. For instance, while the business logic function \textit{generate\_UUID\_300times()} is infrequently sampled by profilers, it calls the library function \textit{UUID.randomUUID()} 300 times, significantly contributing to the overall CPU time consumption of the program.

Flamegraphs \citep{flamegraph} provide a valuable solution to this issue. They effectively illustrate call stacks during the execution of a specific service instance, enabling in-depth analysis of function invocation patterns. Flamegraphs are recognized as powerful tools for examining the performance metrics of individual service instances. However, in the complex ecosystem of a data center, services are often distributed across multiple nodes, resulting in numerous instances of the same service. In such scenarios, the execution patterns of these instances may vary due to diverse user requests and unpredictable factors. Thus, relying solely on the profiling results of a single instance to generate a flamegraph does not accurately reflect the performance characteristics of the entire service.

To tackle this issue, \textbf{Atys} implements a strategy that aggregates raw profile results from all sampled instances of the target service, yielding a comprehensive flamegraph for the entire service. 
This aggregation occurs at two levels.
First, each worker node profiles the target service locally to generate a flamegraph. The flamegraphs from individual worker nodes contain aggregated call stacks, significantly reducing data size compared to the original stack traces. For instance, a raw stack trace file produced by Specjbb2015 at a sampling rate of 10,000Hz may exceed 1GB, while the corresponding aggregated flamegraph is no larger than 10MB, whose data volume is only about 1\% of the former. These pre-processed flamegraphs are then sent to the controller. 
Second, the controller extracts function invocation relationships and metrics corresponding to each function from  received flamegraphs. It then merges the metrics of functions with the same name located at the same level of the invocation chain in all instances, while preserving the original invocation relationships between the caller and called functions. This process employs these consolidated stack traces to produce a comprehensive flamegraph.

This two-level method significantly reduces 99\% data transmission overhead and alleviates the computational load on the controller. In the case of too many instances, in order to further alleviate the data transmission pressure on the controller, it is also possible to first group local nodes and select a central node to aggregate flamegraph within each group. Repeating this process at multiple levels, the controller is only responsible for the final aggregation and presenting results to the user.

\begin{figure*}
\vspace{-10pt}
\centering
\subfloat[Flamegraph of instance 1]{
\includegraphics[width=1\textwidth]{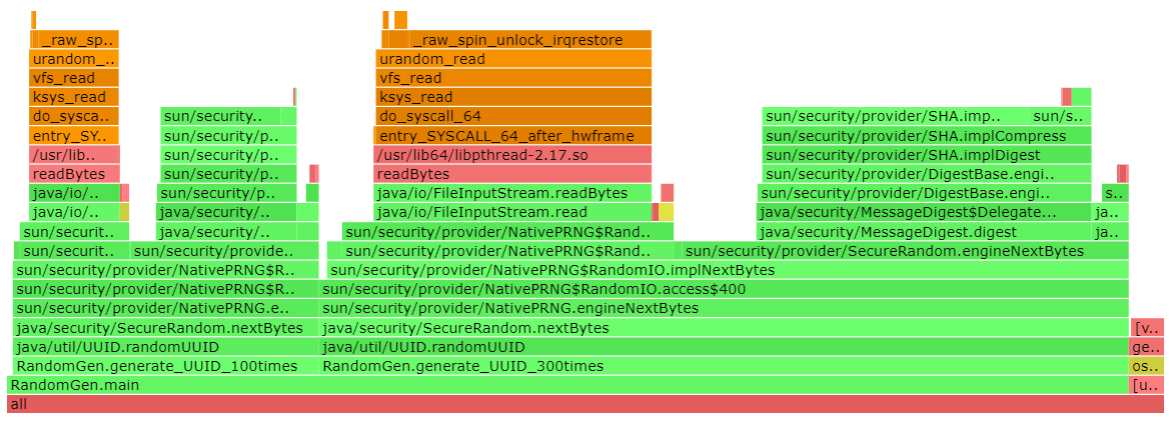}
}
\hfill
\subfloat[Flamegraph of instance 2]{
\includegraphics[width=1\textwidth]{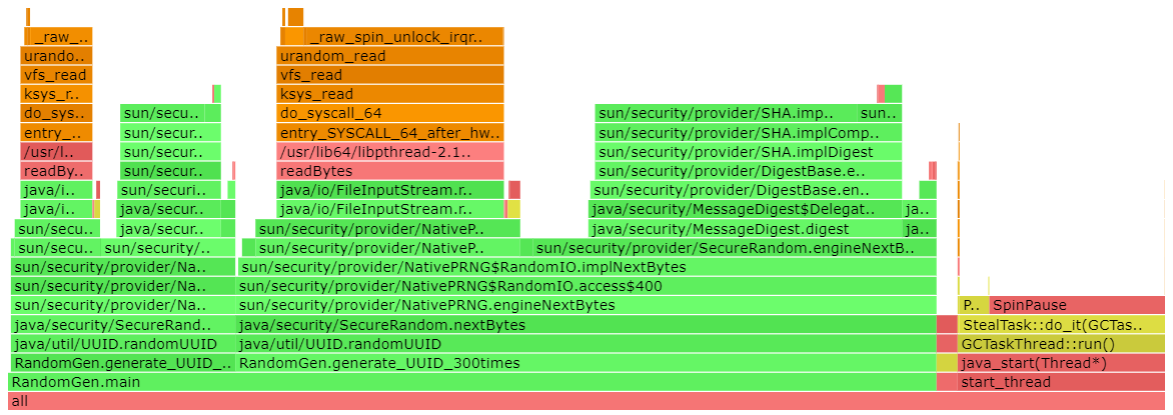}
}
\hfill
\subfloat[Aggregated flamegraph]
{
\includegraphics[width=1\textwidth]{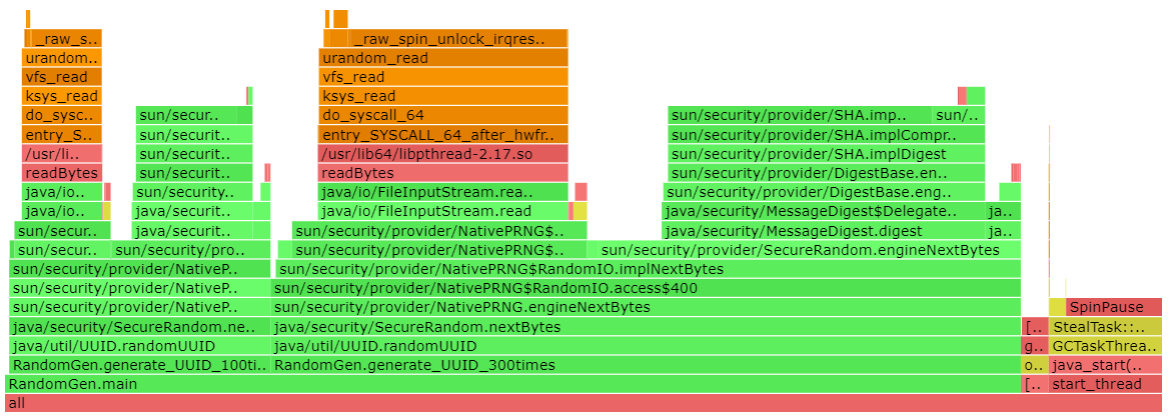}%
}
\caption{Flamegraph of two instances of \textit{RandomGen} and the aggregated flamegraph}
\label{random_agg}
\vspace{-10pt}
\end{figure*}

In the example above, we can better simulate a real-world execution pattern by initiating user requests that randomly  invoke \textit{generate\_UUID\_100times()} and \textit{generate\_UUID\_300times()} in a Poisson distribution.
Figure \ref{random1} and Figure \ref{random2} represent the results of profiling two distinct instances of the modified \textit{RandomGen} program. The differences observed in these two flamegraphs can be attributed to the random factors introduced in the program. For instance, Figure \ref{random1} barely includes the function \textit{spinpause()}. However, in Figure \ref{random2}, \textit{spinpause()} consumes 15.24\% of the CPU time.
Figure \ref{random_overall} presents the aggregated flamegraph, which combines the call stacks from both instances. This comprehensive summary provides a more precise depiction of the service's performance, demonstrating that \textit{spinpause()} accounts for 8.36\% of the CPU usage across all instances. 
This aggregation helps to smooth out variances caused by randomness and provides a clearer picture of the service's true hotspots, which is valuable for performance analysis and optimization.

\section{Efficiency Optimization}
\label{eff-opt}
As noted in Section \ref{sec:intro}, in large-scale distributed microservices, even minor cost reductions in collecting and aggregating stack traces can lead to substantial savings throughout the system. To achieve this, we optimize the efficiency of \textbf{Atys} based on two key observations.

\textbf{Existence of Massive Daemon Threads:} Stack traces from numerous daemon threads often lack informative data for pinpointing hotspot functions, as these threads typically handle background tasks unrelated to performance bottlenecks. By eliminating stack traces from these daemon threads, we can improve the efficiency of the aggregation process.

\textbf{Periodic Variation of Hotspot Functions:} Previous studies have shown that microservice workloads display periodic variation in user requests \citep{KAEinformer}. This is further illustrated by the Java server benchmark Specjbb2015 \citep{specjbb}. By recognizing and leveraging this periodicity, we can dynamically adjust stack trace sampling frequency, resulting in reduced sampling costs while enhancing accuracy.

\subsection{Function Selective Pruning (FSP) Strategy}
\label{sec: pruning}
In this subsection, we demonstrate that it is unnecessary to consider every stack trace when identifying hotspot functions in a software program. We introduce an efficient method for selectively pruning less relevant function calls. This strategy significantly reduces the computational overhead associated with pinpointing hotspot functions while maintaining the accuracy of the results.

\begin{figure}[tbp]
\centering
\includegraphics[width=0.5\textwidth]{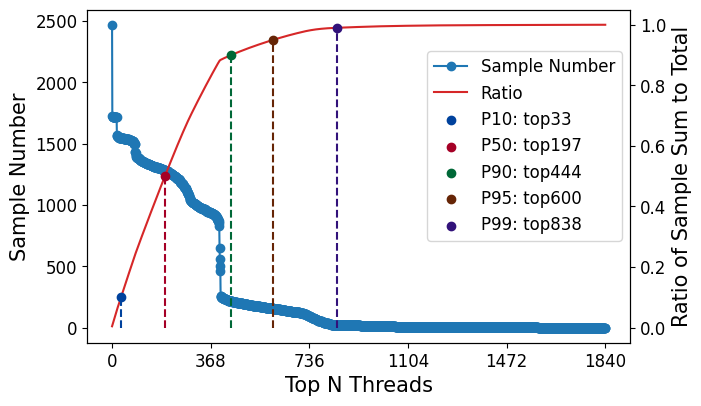}
\vspace{-3mm}
\caption{Number of samples and the ratio of sample sum to total for each subthread in Specjbb2015.}
\vspace{-3mm}
\label{CDF-and-samsum}
\end{figure}

In complex production environments with high concurrency, the target service initiates a significant number of threads \citep{7372280}. A pertinent example is the Java benchmark Specjbb2015 \citep{specjbb}. Figure \ref{CDF-and-samsum} illustrates the distribution of sample counts across various threads during a specific run. The threads are ordered by descending sample counts, with their ranking based on sample volume displayed on the horizontal axis. The left vertical axis represents the number of samples per thread, while the right vertical axis indicates the cumulative percentage of samples contributed by the top n threads relative to the total sample count.
The data indicate a notable concentration of activity among a small subset of threads. Specifically, approximately 25\% of the threads—namely, the top 444 out of 1,840—account for 90\% of the samples, defining the 90th percentile or P90 threads. The majority of threads outside this core group remain inactive, often functioning as daemon threads. Furthermore, the P99 threads, comprising the top 838 threads, demonstrate that the remaining 1,002 threads contribute only 1\% to the total sample count.

In multithreaded services, analyzing the performance of individual threads is crucial for understanding the service's operational dynamics. However, examining stack traces from a single thread does not provide a complete picture of the overall performance landscape. Thus, it is essential to combine stack trace data from all threads to gain a comprehensive view of performance hotspots.
Although it may seem thorough to include every stack trace in the aggregation to avoid overlooking relevant data, this indiscriminate approach can increase computational overhead. More importantly, it risks cluttering the analysis with irrelevant information, obscuring the actual performance bottlenecks.
Consequently, prior to aggregation, it is vital to carefully assess the significance of each stack trace.
To enhance the efficiency and effectiveness of the aggregation process, it is recommended to filter out stack traces that contribute minimally to identifying hotspot functions. This pruning is executed by the local profiler in \textbf{Atys}, which aggregates only the significant raw call stacks.

In \textbf{Atys}, the agent captures stack traces from threads and organizes them by the number of samples per thread. Given that the number of threads is significantly lower than the number of stack traces, the sorting cost is minimal. The agent retains threads with the highest sample counts, such as P99 threads, while discarding samples mainly from inactive daemon threads. This method enables focused analysis of the most active threads, improving profiling efficiency without being burdened by less relevant data. Experiments will demonstrate that this approach yields negligible error and substantial cost savings.

\subsection{Frequency Dynamic Adjustment (FDA) Scheme}

Through experiments and data analysis, we have made two key observations regarding the relationship between sampling frequency and accuracy, as well as the periodic variation of hotspot functions. 

Firstly, profiling at a higher sampling frequency is widely recognized as a means to achieve greater accuracy due to the increased number of samples. However, this approach incurs higher costs, provided the frequency does not surpass a certain threshold. For individual programs, users may opt for a higher sampling frequency, sacrificing some performance for enhanced profiling accuracy. Yet, in the context of multiple instances, this may not be a viable solution. Consequently, a trade-off should be made between cost and accuracy.

\begin{figure}[tb]
    \centering
    \subfloat[Specjbb2015]{
    \includegraphics[width=0.45\textwidth]{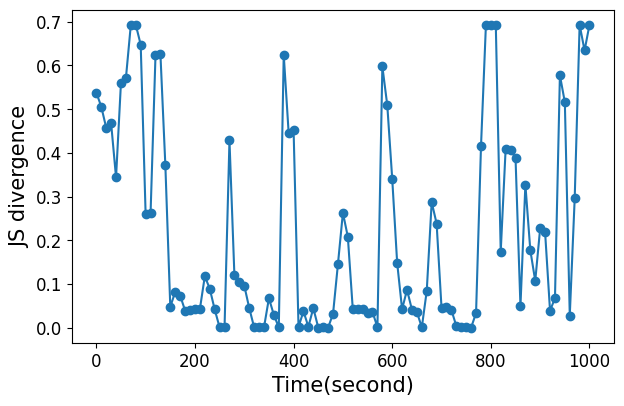}
    }
   \hfill
    \subfloat[VGG16 inference]{
    \includegraphics[width=0.45\textwidth]{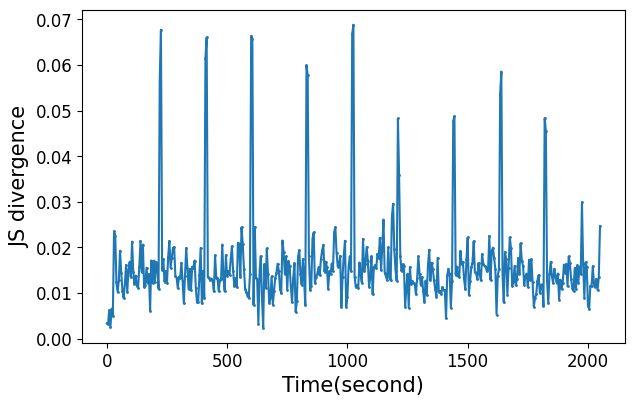}
    }
    \caption{The periodic variation of hotspot functions in typical workloads}
    \label{hotspot-shift}
    \vspace{-3mm}
\end{figure}

Secondly, our research reveals that certain workloads exhibit periodic fluctuations in their hotspot functions. To investigate this pattern, we conducted experiments using two benchmarks: the widely recognized Java benchmark Specjbb2015 \citep{specjbb} and a representative deep learning inference task employing the VGG16 architecture \citep{Simonyan2014VeryDC}, which iteratively processes the CIFAR dataset \citep{Krizhevsky2009LearningML}. We meticulously monitored the variations in the top 10 hotspot functions during the execution of these benchmarks.
We specifically analyze the proportion of time spent on the top 10 hotspot functions as a probability distribution. By employing the $Jenson-Shannon (JS)$ divergence \citep{jsdiv}, we quantify the differences between the current and previous distributions at 10-second intervals. The findings, illustrated in Figure \ref{hotspot-shift}, highlight two significant trends:
1) High levels of JS divergence are consistently observed at the beginning and end of the application's lifecycle, indicating rapid changes in hotspot functions during these phases.
2) During stable application operation, a cyclical variation pattern in hotspot functions emerges, characterized by recurring spikes in JS divergence at regular intervals, reflecting shifts in hotspots that subsequently return to lower levels before the next cycle commences.

Motivated by these observations, we propose a dynamic frequency adjustment scheme that adaptively modifies the sampling rate based on profiling results. The goal is to minimize profiling costs while maintaining accuracy. The fundamental idea is to increase the sampling frequency when significant changes in hotspots are detected, and to either maintain or decrease it when they stabilize, thereby improving the accuracy of low-frequency sampling results with high-frequency samples.

The principle behind this approach can be outlined as follows: leveraging high-frequency sampling results as reference points to enhance the accuracy of low-frequency sampling outcomes. 
Initially, we gather precise results through high sampling rates, followed by obtaining data at lower sampling rates. If the discrepancy between low and high-frequency sampling results is minimal, it indicates that the hotspot functions have remained relatively stable. Consequently, the high-frequency results can substitute the low-frequency sampling outcomes. Conversely, significant differences suggest that the hotspot functions are changing rapidly, rendering previously obtained high-frequency results no longer applicable. In such instances, the sampling rate should be increased to acquire new, accurate reference points.

The frequency in \textbf{Atys} is adjusted exponentially, where the new frequency is determined by multiplying the current frequency by a user-specified decaying coefficient $\lambda$. To assess the similarity of the sampling results, we employ the $JS$ divergence of the distributions of hotspot functions. Specifically, we treat the percentage of time each function occupies in the results of two consecutive samples as two probability distributions. Thus, ${P}_i = \{f_{i,1}, f_{i,2}, \ldots, f_{i,n}\}$, where $P_i$ denotes the distribution obtained from the $i$th sampling, and $f_{i,k}$ indicates the proportion of time that the $k$th function occupies across all samples collected at the $i$th instance. We can then compute the divergence between two distributions. Two distributions are considered significantly different when the condition $D_{JS} > \theta$ is satisfied, where $\theta$ is a user-defined threshold determining the strictness of frequency adjustment. The pseudo-code of this approach is presented in Algorithm \ref{dyn freq alg}. 

\begin{algorithm} [tb]
	\caption{Dynamic Frequency Adjustment} 
	\label{dyn freq alg} 
	\begin{algorithmic}
        \STATE \textbf{For each time window:}
		\REQUIRE $\theta, \lambda$~\COMMENT{$\theta, \lambda \in (0,1)$}, $f$: current time window's frequency, $R_0$: profiling result of last time window, $R$: profiling result of current time window
        \ENSURE $f'$: sampling frequency for next time window
            \STATE $P_0 \gets \text{Distribution of functions in } R_0$
            \STATE $P \gets \text{Distribution of functions in } R$
            \STATE $D \gets \text{JS divergence}(P,P')$
            \IF{$D > \theta$}
                \STATE $f' = f/\lambda$
            \ELSIF{$D \leq \theta $
             in more than the specified number of consecutive newline time windows\footnotemark}
                \STATE $f' \gets f*\lambda$
            \ELSE 
                \STATE $f' \gets f$
            \ENDIF
            \RETURN $f'$
	\end{algorithmic} 
\end{algorithm}
\footnotetext[3]{This threshold is adaptable based on user preferences. For our experiment, we set it to 5.}

To adjust the strictness of similarity evaluation, users can set the threshold $\theta$. 
The optimal value of $\theta$ may vary across different programs. To establish an initial suitable value, users can perform a initialization stage by a low-cost, coarse, and offline profiling before large-scale profiling. This determined value can subsequently be applied directly, incurring no additional costs during ongoing monitoring activities.

To reduce computational costs, not all functions in the sampling results are considered when calculating $D_{JS}$, due to the typically large number of functions. In most workloads, the top 10 hotspot functions account for a significant portion of the total cost, highlighting the primary performance bottlenecks of the target service. In our specific implementation, we demonstrate the use of the top 10 hotspot functions for calculating $D_{JS}$, though this selection is flexible. Users can adjust this parameter; increasing its value incorporates more functions, enhancing accuracy in identifying changes in hotspot functions, albeit at a higher computational cost.

\subsection{Initialization Stage}
The essential purpose of both \textit{Function Selective Pruning} and \textit{ Frequency Dynamic Adjustment} is to balance cost and accuracy, which requires knowing how they vary according to different parameters. Moreover, different services have different performance characteristics and access patterns, and the profiling cost and accuracy of profiling kernels themselves are also different. It is impossible to know in advance the effects of different parameters before profiling in practice. Therefore, to determine the optimal parameters for them, we need an initialization stage by performing a pre-profiling before large-scale profiling.
Since our target is long-running large-scale distributed services, the cost of this one-time initialization process is usually acceptable compared to the cost of subsequent long-term monitoring.

For the pruning standards of \textit{Function Selective Pruning}, there is a nearly functional relationship between the percentage of pruned threads, the time reduced, and the mean average percentage error(MAPE) caused for the target service, as shown in Figure \ref{prune_eval}. We can fit this relationship through the results of pre-profiling and appropriate models. Therefore, given the acceptable maximum MAPE, we can calculate the maximum pruning percentage to achieve the maximum time reduction within the error range. This can be formalized as the following optimization problem:
\begin{equation}
\begin{array}{ccclcl}
\displaystyle \min_{p} & T(p) \\
\textrm{s.t.} & MAPE(p) \leq \epsilon\\
\end{array}
\end{equation}
where $p$ is the pruning percentage, while $\epsilon$ denotes the maximum acceptable MAPE. The functions $T(p)$ and $MAPE(p)$ illustrate the fitted relationships between $p$ and the corresponding time reduction, as well as the MAPE induced by $p$, respectively.
Since $T(p)$ and $MAPE(p)$ are both monotonically increasing, this problem can be easily solved.

For the example Specjbb service, we choose linear model for $T(p)$, and logarithmic model for $MAPE(p)$ (MAPE of the top50 hotspot functions). The results are $T(p)=-1.0614p + 114.44$, fitting MAPE=$4.75$; and $MAPE(p)=984.368 * log{(-0.001p+1.099)}$, fitting MAPE=$15.73$.
It should be noted that the MAPE of the logarithmic model before P99 is only 1.66, and at P99, the overall relative error increases due to the small absolute value of the error. This means that the predicted values are highly accurate in most feasible regions of this optimization problem.
Given a threshold $\epsilon$ in the example, we can thus calculate $p\geq \frac{e^{\epsilon/984.368}-1.099}{-0.001}$.

As for the determination of the threshold of Frequency Dynamic Adjustment Scheme, 
user can derive an image of JS divergence over time like Figure \ref{hotspot-shift} through pre-profiling. Through this image, user can easily see the migration of hot functions of the service, and thus set an appropriate JS divergence threshold for frequency adjustment. The general target is to make the small changes in the hotspot function of the service do not reach the threshold most of the time, while those drastic changes reach it.

\section{Evaluation}
\label{eval}
\subsection{Experiment Setup}
We conduct experiments on a 13-node cluster to evaluate the effectiveness of function selective pruning and frequency dynamic adjustment strategies, along with the scalability of \textbf{Atys}. Specifically, we utilize SPECjbb2015 \citep{specjbb} as the Java benchmark and async-profiler \citep{asyncprof} as the Java profiling kernel. Additionally, we employ the VGG16 network inference \citep{Simonyan2014VeryDC} for Python benchmarking, using py-spy \citep{pyspy} as the profiling kernel.

The Specjbb2015 runs in HBIR (High Bound Injection Rate) mode, indicating it seeks HBIR before exiting. The VGG workload is set up to employ the VGG16 network for repeated classification on the CIFAR-10 dataset \citep{Krizhevsky2009LearningML}. All benchmark programs are deployed via Kubernetes \citep{k8s}, with \textbf{Atys} installed as a daemon on the target node. All tests are performed on identical worker nodes, as detailed in Table \ref{node_config}.

\subsection{Baselines}
Our advancements in profiling kernel efficiency enhance the kernels' optimizations. This indicates that the optimization techniques outlined in \textbf{Atys} can be applied to both single-machine and multi-machine profilers. 
Furthermore, due to the inherent performance disparities among different kernels, it would be inappropriate to evaluate distinct kernels using our optimization techniques. Consequently, our experiments concentrate on assessing the optimizations on the same kernel, rather than merely deploying the kernel from previous studies. We select the open-source multi-machine profiler pyroscope \citep{pyroscope} as the baseline, which employs the same Java and Python profiling kernels as \textbf{Atys}.

\begin{table}[tbp]
\caption{Configuration of worker nodes in the experiments}
\vspace{-3mm}
\label{node_config}
\centering
\begin{small}
\begin{tabular}{|m{0.3\columnwidth}|m{0.6\columnwidth}|}
\hline
\centering OS  & \centering Alibaba Cloud Linux release 2.1903 LTS \tabularnewline \hline
\centering CPU & \centering Intel(R) Xeon(R) Platinum 8269CY CPU @ 2.50GHz \tabularnewline \hline
\centering Core Number   & \centering 104 \tabularnewline \hline
\centering Memory Size   & \centering 187GB \tabularnewline \hline
\end{tabular}
\end{small}
\end{table}

\begin{figure}[t]
\centering
\includegraphics[width=0.5\textwidth]{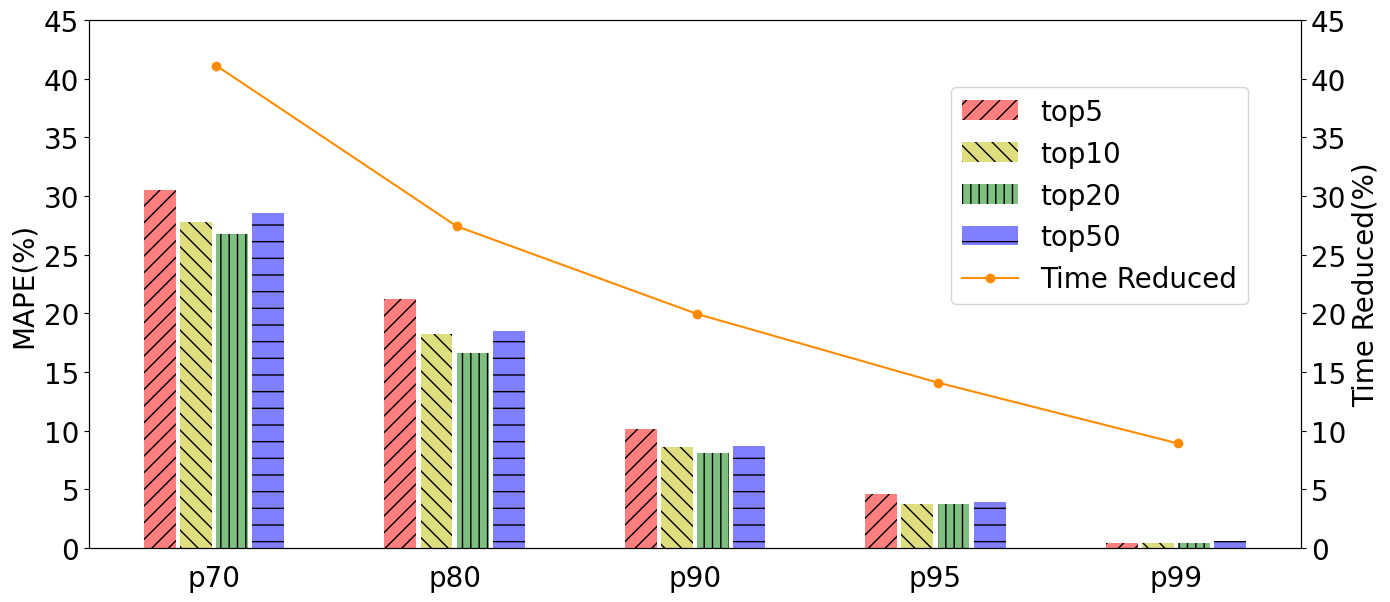}
\vspace{-2mm}
\caption{MAPE of the top n hotspot functions and the percentage of time saved under various pruning configurations.}
\label{prune_eval}
\end{figure}

\subsection{Effect of FSP Strategy}
We employ the function selective pruning (FSP) strategy, as detailed in Section \ref{sec: pruning}, on the stack traces gathered from a specific execution of Specjbb2015. We then analyze the mean absolute percentage error (MAPE) of the execution time for the top n most sampled functions, alongside the aggregation time under varying pruning thresholds, with n set to 5, 10, 20, and 50. Notably, the time to sort 1,840 threads by sample count is minimal compared to the aggregation cost of hundreds of MBs of raw stack traces. Therefore, our primary focus is on the aggregation process's time cost for comparison.

The results are depicted in Figure \ref{prune_eval}, showing that for the top 5, 10, 20, and 50 hotspot functions, the FSP strategy in \textbf{Atys} yields significant time savings, surpassing the marginal error incurred. Specifically, with the P99 pruning threshold, a mere 0.58\% MAPE is introduced for the top 50 hotspot functions, while aggregation time decreases by 6.8\%. Therefore, the P99 standard is established as the default pruning threshold in \textbf{Atys}.

\begin{figure*}[!t]
    \centering
    \subfloat[VGG16 inference]{
    \includegraphics[width=0.48\textwidth]{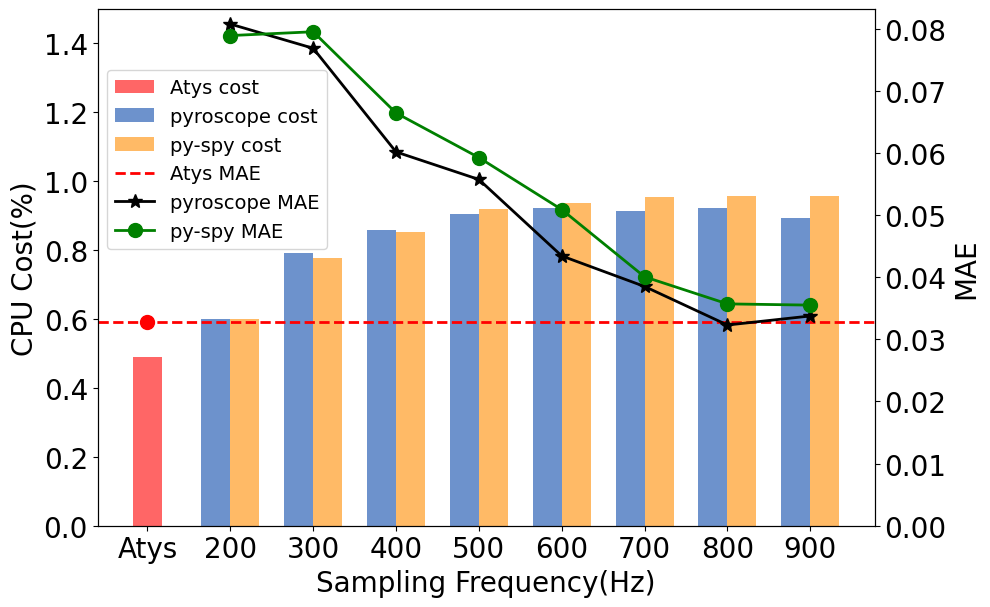}
            }
    \subfloat[Specjbb2015]{
    \includegraphics[width=0.48\textwidth]{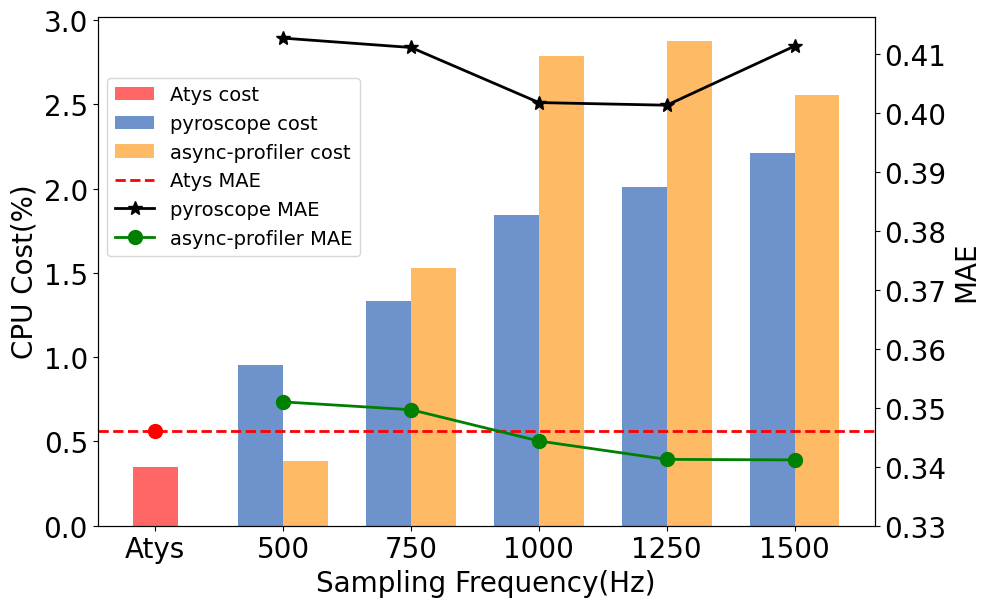}
    }
    \vspace{-0mm}
    \caption{CPU cost and MAE with different sampling frequencies by \textbf{Atys}, single-machine profilers, and pyroscope respectively}
    \label{cost_acc_dyn}
    \vspace{-0mm}
\end{figure*}

\subsection{Effect of FDA Scheme}
To evaluate the effectiveness of the frequency dynamic adjustment (FDA) scheme, we initially select six time-consuming ($>5\%$) business logic functions from Specjbb2015, as detailed below:
\begin{tcolorbox}
\begin{itemize}
    \item AssociativityOfCategoryTransaction.execute 
    \item SupermarketAudit.analyze
    \item SupermarketAudit.summarizeReceipt
    \item InStorePurchaseTransaction.doCheckout
    \item InStorePurchaseTransaction.doPurchase
    \item OnlinePurchaseTransaction.doCheckout
\end{itemize}
\end{tcolorbox}
\noindent In addition, we select four functions essential for neural network inference from the VGG16 benchmark:
\begin{tcolorbox}
\begin{itemize}
    \item forward (torch/nn/modules/linear.py)
    \item \_conv\_forward (torch/nn/modules/conv.py)
    \item \_max\_pool2d (torch/nn/functional.py)
    \item relu (torch/nn/functional.py)
\end{itemize}
\end{tcolorbox}

We have instrumented the benchmark code to accurately measure the CPU time consumed by each function. For the Java benchmark, we achieve this by invoking \textit{threadMXBean.getCurrent} \textit{ThreadCpuTime()} before and after the function's execution. In the case of the Python benchmark, we employ the \textit{CProfile} \citep{cprof} library. Furthermore, we extrapolate the CPU time based on profiling results. The percentage of time consumed by these hotspot functions is then converted into a distribution. Finally, we compute the difference between the distribution derived from profiling and the actual time consumption, expressed in terms of mean average error (MAE).

We calculate the MAE and CPU consumption of profiling across various sampling frequencies using different profilers. The parameters for the FDA scheme are set as follows: the JS divergence threshold $\theta_{specjbb}$ = $0.5$, $\theta_{vgg}$ = $0.05$, and the decay rate $\lambda$ = $0.8$.

Several observations can be made from the results depicted in Figure \ref{cost_acc_dyn}. Firstly, as the sampling frequency increases, sampling costs generally rise while sampling error decreases, confirming the trend discussed in Section \ref{eff-opt}. 
Secondly, in the Python benchmark, \textbf{Atys} consistently achieves lower error and costs than both py-spy and pyroscope across all tested frequencies using our FDA strategy. Its MAE approaches that of pyroscope at 800Hz, with CPU utilization at only 53\% of pyroscope's.
Thirdly, in the Java benchmark, \textbf{Atys} has an average sampling frequency of 357.8Hz, with CPU usage comparable to async-profiler at 5,00Hz. However, it demonstrates a significantly lower MAE than async-profiler at this frequency, nearing the performance at 1,000Hz while incurring only 12.4\% of the sampling cost. Additionally, \textbf{Atys}' MAE remains notably lower than that of pyroscope across all frequencies.
These results demonstrate the effectiveness of \textbf{Atys}'s FDA strategy, achieving high accuracy while significantly minimizing performance overhead.

\begin{figure*}[t]
    \centering
    \subfloat[Memory usage]{
    \includegraphics[width=0.8\textwidth]{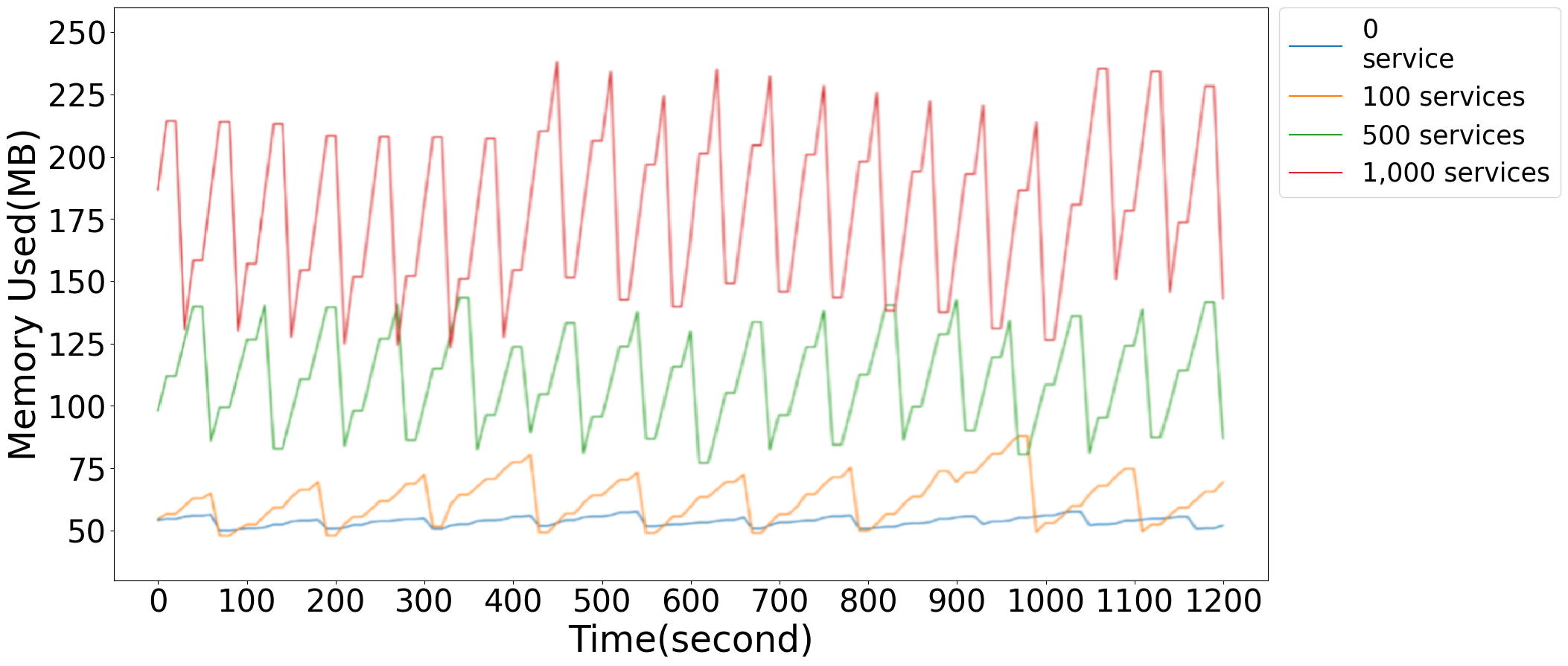}
    }
    \hfill
    \subfloat[CPU usage]{
    \includegraphics[width=0.8\textwidth]{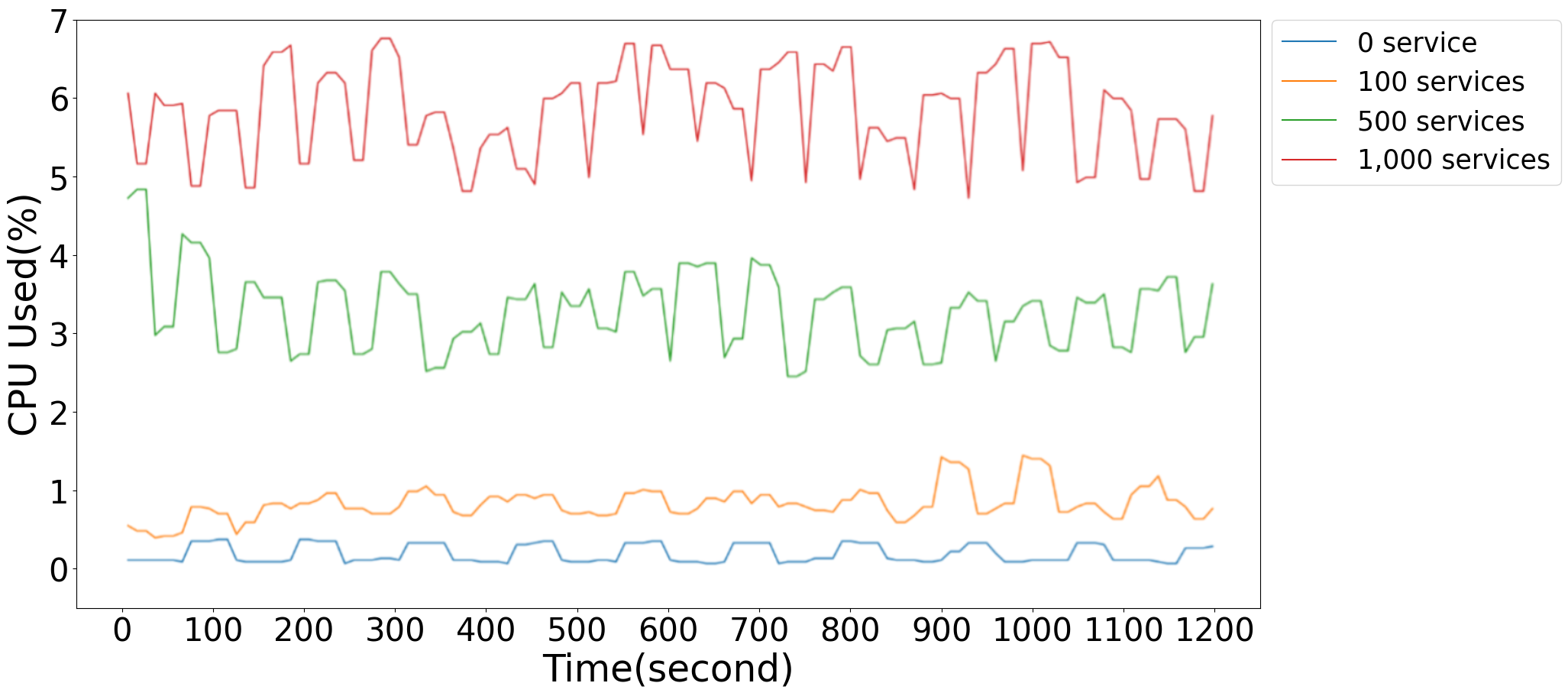}
            }
    \vspace{-0mm}
    \caption{ Memory and CPU usage of the Prometheus server with varying numbers of monitoring instances.}
    \label{pm-eval}
    \vspace{-0mm}
\end{figure*}

\subsection{System Scalability}
When deploying \textbf{Atys} in production environments, the Prometheus server must track and store extensive metrics, potentially resulting in significant strain. Thus, evaluating the server's capacity to manage increased workloads during simultaneous monitoring of multiple instances is crucial.
We conduct experiments in which \textbf{Atys} concurrently monitors 0, 100, 500, and 1,000 instances of web services over a period of 1,200 seconds. For each instance, \textbf{Atys} is configured to collect the top 10 hotspot functions. The memory consumption of the Prometheus service is recorded and presented in Figure \ref{pm-mem}. 

Initially, with no active monitoring, the Prometheus server's memory usage is approximately 50MB. As the number of monitored services increases, memory consumption rises accordingly. However, even when monitoring up to 1,000 service instances, the memory footprint remains below 250MB, which is negligible compared to the worker node's total memory of 184GB. 
The CPU usage of the Prometheus service is also monitored, as illustrated in Figure \ref{pm-cpu}. At its peak workload, while monitoring 1,000 service instances simultaneously, the server utilizes only about 6\% of the host node's total CPU.

These results confirm that the Prometheus server has ample capacity to manage numerous sampling tasks without strain. Furthermore, this demonstration of the server's resource efficiency underscores the suitability of \textbf{Atys} for meeting the demands of extensive profiling tasks in production large-scale microservice environments.

\section{Conclusion}
This work presents \textbf{Atys}, a highly efficient profiling framework tailored for the complexities of large-scale services in production environments. \textbf{Atys} incorporates a language-agnostic adaptation mechanism, ensuring compatibility with various programming languages, both compiled and interpreted. Additionally, it offers a flamegraph aggregation method for a comprehensive view across multiple service instances.
To enhance efficiency, \textbf{Atys} implements a function selective pruning strategy and a frequency dynamic adjustment scheme to minimize profiling overhead. The performance and effectiveness of \textbf{Atys} have been validated through a prototype, which underwent extensive experiments, showcasing its potential for practical deployment in production environments.

\bibliographystyle{elsarticle-num} 
\bibliography{ref}

\end{document}